\documentclass[twocolumn,nofootinbib,showpacs,prd,superscriptaddress]{revtex4}
\usepackage{amsmath}
\usepackage{amsfonts}
\usepackage{amssymb}
\usepackage{graphicx}
\usepackage{epstopdf}
\usepackage{subfigure}

\begin{document}

\title{A numerical approach to model independently reconstruct $f(R)$ functions through cosmographic data}

\author{Liberato Pizza}
\affiliation{Dipartimento di Fisica, Universit\`a di Pisa, Largo Bruno Pontecorvo, i-56127, Pisa, Italy.}
\affiliation{Istituto Nazionale di Fisica Nucleare (INFN), Sezione di Pisa, Largo Bruno Pontecorvo, i-56127, Pisa, Italy.}

\begin{abstract}
The challenging issue of determining the correct $f(R)$ among several possibilities is here revised by means of numerical reconstructions of the modified Friedmann equations around the redshift interval $z\in[0,1]$. Frequently, a severe degeneracy between $f(R)$ approaches occurs, since different paradigms correctly explain present time dynamics.   To set the initial conditions on the $f(R)$ functions, we involve the use of the so called cosmography of the Universe, i.e. the technique of fixing constraints on the observable Universe by comparing expanded observables with current data. This powerful approach is essentially model independent and correspondingly we got a model independent reconstruction of $f(R(z))$ classes within the interval $z\in[0,1]$. To allow the Hubble rate to evolve around $z\leq1$, we considered three relevant frameworks of effective cosmological dynamics, i.e. the $\Lambda$CDM model, the CPL parametrization and a polynomial approach to dark energy. Finally cumbersome algebra permits  to pass from $f(z)$ to $f(R)$ and  the general outcome of our work is the determination of a viable $f(R)$ function, that  effectively describes the observed Universe dynamics.
\end{abstract}


\maketitle

%%%%%%%%%%%%%%%%%%%%%%%%%%%%%%%%%%%%%%%%%%%%%%%%%%%%%%
\section{Introduction}
%%%%%%%%%%%%%%%%%%%%%%%%%%%%%%%%%%%%%%%%%%%%%%%%%%%%%%

A current speeding up Universe was discovered almost two decades ago, by using 42 type Ia supernovae as standard indicators \cite{nobel,nonobel,nonobel2}. Further data have definitively forecasted this experimental evidence, confirming that the Universe is accelerated at late times \cite{noleb}. Particularly, after a particular transition epoch \cite{transition}, the Universe started unexpectedly to accelerate, showing that only standard matter cannot be responsible for its present dynamics \cite{review}. The most accepted interpretation includes dark matter and dark energy, respectively responsible for structure formation and repulsive dynamics \cite{review3}. In particular, the first component, i.e. dark matter, permits structures to cluster and form during early phases of Universe's evolution, whereas dark energy seems to influence current dynamics, enabling the Universe to positively accelerate at small redshift  \cite{darkmedarke}. The most accredited paradigm is named \emph{the $\Lambda$CDM model} and assumes a non-evolving vacuum energy cosmological constant $\Lambda$, which dominates over baryons and dark matter at our epoch \cite{Lambda,LambdaFit}. Moreover, present data also prospect a spatially flat Universe, characterized by baryons and cold dark matter which accounts for almost the $25\%$ of the whole energy content, while the $\Lambda$ density for about the $68\%$ \cite{planck}. Quite surprisingly, the corresponding dark energy equation of state for the pressure is negative, leading to a non-clear physical interpretation of the main components of the energy momentum tensor \cite{ecco}. Thus, the observed speed up description is so far theoretically uncomplete and leads cosmologists to suppose that general relativity breaks down at precise energy scales \cite{bamba}. This fact would provide the expected repulsive effects, capable of accelerating the Universe today \cite{review2}. Hence, a self consistent enlargement of general relativity reviews both dark matter and energy effects as fundamental properties of a single theory. Those frameworks represent the well-known \emph{modified theories of gravity}, consisting of many models trying to include higher curvature terms, the torsion scalar, additional scalar fields, novel particles derived from supersymmetry and so forth (see \cite{reviewona} for details). Among several possibilities, a feasible extension of general relativity, which reproduces the dark energy effects, is the class of $f(R)$ models. Those paradigms manage to replace the Ricci scalar $R$ in terms of a generic analytic function $f(R)$, modifying correspondingly the Einstein-Hilbert action $S$ as
\begin{equation} \label{actionmetric}
S=\frac{1}{2\kappa} \int d^4x \,
\sqrt{-g}\, f(R)+S^{(m)} \;,
\end{equation}
where $\kappa$ is a coupling constant related to the gravitational constant $G$, $g$ the spacetime involved and $S^{(m)}$ the matter action. Since $f(R)$ is not known \emph{a priori}, there exist many formulations of modified $f(R)$ theories \cite{formulations} passing almost all the available experimental tests at the solar system regime. Hence, there exists a strong degeneracy among cosmological models derived from $f(R)$ approaches and the correct form of the $f(R)$ function is still object of debate. However, any proposed $f(R)$ represents a \emph{ad hoc} formulation for describing the Universe dynamics \cite{bariolepto}. Frequently, in fact, the particular choice of $f(R)$  is physically unmotivated and so the need of puzzling physical $f(R)$ represents the main issue related to the $f(R)$ picture.

The main purpose of our paper is to discriminate, among different sets of $f(R)$ possibilities, the ones passing the local bounds offered by cosmography and to propose a model independent reconstruction of cosmographic $f(R)$ functions. To perform our numerical analyses, we need to numerically solve the modified Friedmann equations. The modified Friedmann equations explicitly depend upon the Hubble rate $H$ and derivatives, the Ricci scalar $R$ and the $f(R)$ function and derivatives. Hence, to express the Friedmann equations in terms of a single variable, we rewrite them in function of the redshift $z$. To fix the evolution of $H$ and then to  calibrate the corresponding Hubble rates, entering the Friedmann equations, we assume that the curvature dark energy fluid is mimicked in terms of three consolidate dark energy candidates. In particular, we approximate the Hubble evolution using the $\Lambda$CDM model, the Chevallier-Polarski-Linder (CPL)  parametrization and finally a phenomenological dark energy approximation (Starobinski \cite{staro}). Afterwards, we reconstruct the behavior of the $f(R)$ function by numerically solving the above cited cosmological equations, fixing the free parameters to agree with the cosmographic bounds. In particular, we consider a numerical approach to determine the behaviors of $f(R)$, passing through the corresponding $f(z)$ function, which is biunivocally determined by considering the dependence upon the redshift of the Ricci scalar $R$. Hence, we rewrite the $f(R)$ in terms of a redshift function only, i.e. $f(z)$, which consists in replacing the time dependence of $R$ by means of the more feasible variable $z$. Employing cosmological data around $z\in[0,1]$, in which the most of data is concentrated, we depict the shapes of $f(z)$, $\rho_{curv}(z)$, $P_{curv}(z)$ and $\omega_{curv}$, i.e. the curvature equation of state of the fluid responsible for the Universe speeding up in function of $z$ only. Afterwards, inverting the numerical equations, i.e. passing from the redshift $z$ to $R$, we will be able to build up the numerical behavior of $f(R)$ in function of $R$ as well. Further, we demonstrate that our method involves the smallest number of impositions possible, since to set \emph{model independent} initial conditions on the involved Friedmann equations, we make use of the above cited  \emph{cosmography} of the Universe. Indeed, cosmography represents a method to describe the Universe dynamics, without using any assumptions \emph{a priori}. The strategy of cosmography precisely assesses numerical settings in the redshift domain $z\in[0,1]$ and it represents an accurate way to bound $H_0$, the Universe acceleration $q$ and its variation $j$ at present time. To permit cosmography to act on the initial settings of the Friedmann equations, we propose the strategy of matching $f(z)$ and derivatives with the cosmographic coefficients. Moreover, we assume that the Ricci scalar and the corresponding derivatives may be related to the cosmographic coefficients as well. Thus, we finally compare the cosmological results with the solar system constraints and we infer a class of $f(R)$ functions which better behave as the Universe expands. As a consequence, we propose new viable $f(R)$ functions, which are not postulated but inferred from numerically solve the Friedmann equations. Our treatment lies on a direct reconstruction built up in a model independent way, by using the Universe cosmography. Once our $f(R)$ function is known, we turn back solving the Friedmann equations, showing the behavior of our proposal in functions of the redshift parameter $z$. Thus, our methodology enables to highlight the correct form of the $f(R)$ function, suggesting which one is really suitable for describing current dynamics.

The paper is thus structured: in Sec. II, we focus on the main features of $f(R)$ cosmology, highlighting the basic requirements that every $f(R)$ paradigm should provide. In Sec. III, we describe the basic demands of cosmography as a tool to fix initial settings on the numerics we are going to show. In particular, we describe the role of every cosmographic coefficients and how cosmography can be framed in the context of $f(R)$ gravity. In Sec. IV, we match cosmography with $f(R)$ theories and we show which coefficients are of particular interest in our procedures, emphasizing the peculiar property of cosmography to fix limits on $f(z)$ model independently. Besides we proceed to numerical reconstruct the $f(z)$ function, calibrating the Hubble rate in the interval $z\in[0,1]$ using three different calibrating $H(z)$, i.e. the $\Lambda$CDM model, the CPL parametrization and a phenomenological reconstruction of dark energy. Once the function is determined, we analyze the principal properties and we discuss their implications in modern cosmology.  In Sec. V, we describe the $f(R)$ function and we calculate the corresponding properties in the time domain. We show that the function is compatible with some other paradigms already presented in the literature, although it actually differs from them for some other properties. Finally, Sec. VI is devoted to conclusion and final perspectives of our work \cite{after}.

%%%%%%%%%%%%%%%%%%%%%%%%%%%%%%%%%%%%%%%%%%%%%%%%%%%%%%
\section{Consequences of $f(R)$ cosmology}
%%%%%%%%%%%%%%%%%%%%%%%%%%%%%%%%%%%%%%%%%%%%%%%%%%%%%%

\noindent To determine the dynamical equations in $f(R)$ cosmology, one varies Eq. (\ref{actionmetric}) with respect to the metric $g_{\mu\nu}$. This leads to fourth order field equations of the form \cite{capozcurv, formulations}
\begin{equation}\label{metf}
f'(R)R_{\mu\nu}-\frac{1}{2}f(R)g_{\mu\nu}-\left[\nabla_\mu\nabla_\nu -g_{\mu\nu}\Box\right] f'(R)= \kappa
\,\tilde{T}_{\mu\nu}^{(m)}\,,
\end{equation}
where, baptizing $S^{(m)}$ the matter action, we write down
\begin{equation}\label{set}
\tilde{T}_{\mu\nu}^{(m)}\equiv\frac{-2}{\sqrt{-g}}\, \frac{\delta S^{(m)}
}{\delta g^{\mu\nu} }\,,
\end{equation}
with $\kappa\equiv\frac{8\pi G}{c^4}$ and it will be imposed equal to 1 hereafter. Equivalently, we explicitly split the matter counterpart from curvature dark energy, i.e. $
G_{\mu \nu} = R_{\mu \nu} -  \frac{1}{2} R g_{\mu
\nu} = T^{(curv)}_{\mu \nu} + T^{(m)}_{\mu \nu}$
where
\begin{eqnarray}
T^{(curv)}_{\mu \nu} & = & \frac{1}{f'(R)} \left \{ g_{\mu \nu} \left [ f(R) - R f'(R) \right ] /2 +\right . \nonumber \\
~ & ~ & \nonumber \\
~ & + & \left . f'(R)^{; \alpha \beta} \left ( g_{\alpha \mu} g_{\beta
\nu} - g_{\alpha \beta} g_{\mu \nu} \right ) \right \}\,, \label{eq:curvstress}
\end{eqnarray}
the tensor responsible for curvature corrections, usually named the \emph{curvature energy-momentum tensor}. Note that the prime, i.e. $"{'}"$, denotes the derivative with respect to $R$, while the semicolon, i.e. $"^{;}"$, denotes the covariant derivative. This term represents a source for the whole energy-momentum tensor and permits to reproduce the Universe dynamics by means of curvature corrections. Clearly, by looking at Eq. (\ref{set}), the matter tensor is coupled to the curvature tensor as $T^{(m)}_{\mu \nu} = \tilde{T}^{(m)}_{\mu \nu}/f'(R)$ and the corresponding modified Friedmann equations become \cite{formulations, capozcurv}
\begin{equation}\label{eq: fried1}
H^2  = \frac{1}{3} \left [ \rho_{curv} +
\frac{\rho_m}{f'(R)} \right ]\,,
\end{equation}
and
\begin{equation}\label{eq: fried2}
2\dot{H} +3 H^2 = - P_{curv}- P_{m}\,.
\end{equation}
In these equations we neglected scalar curvature terms, i.e.  $k=0$, according to recent observations \cite{plankkk} and we define the curvature density as
\begin{equation}\label{eq: rhocurv}
\rho_{curv} =   \frac{1}{2} \left( \frac{f(R)}{f'(R)}  - R \right) - 3 H \dot{R} \frac{f''(R)}{f'(R)} \,,
\end{equation}
 the barotropic pressure as
\begin{equation}
P_{curv} = \omega_{curv} \rho_{curv} \label{eq: pcurv}\,,
\end{equation}
and we consider $P_m=0$ because we study Universe expansion in the matter dominated phase. 
For our purposes, we assume the pressure to be barotropic since it does not explicitly depend upon the Universe entropy $S$ \cite{kunzo}. The well-known effective curvature barotropic factor is therefore given by
\begin{equation}
\omega_{curv} = -1 + \frac{\ddot{R} f''(R) + \dot{R} \left [ \dot{R}
f'''(R) - H f''(R) \right ]} {\left [ f(R) - R f'(R) \right ]/2 - 3
H \dot{R} f''(R)}\,, \label{eq: wcurv}
\end{equation}
and represents the effective equation of state for the curvature term.  Finally, the corresponding Ricci scalar $R$ can be framed in terms of the Hubble parameter as \cite{formulations}
\begin{equation}
R = -6 \left ( \dot{H} + 2 H^2\right )\,,
\label{eq: constr}
\end{equation}
where we assume $f^{''}(R)\neq 0$, for $R<R_0$ to hold. In our calculations, $R_0$ represents the Ricci scalar at our time $t_0$, which corresponds to the redshift $z=0$. Later on, we take into account the above cosmological equations and we describe their evolutions by numerically solving Eq. (\ref{eq: fried1}). In particular,  numerical outcomes derived from Eq. (\ref{eq: fried1}) enable us to frame the $f(R)$ corrections and to obtain viable $f(R)$ candidates, which would represent effective classes of cosmological $f(R)$ models. The problem of fixing the initial conditions to use is here overcome by the use of cosmography. In the next paragraph, we report the basic demands offered by cosmography to set the initial conditions on $f(R)$ functions, showing that it is possible to rewrite $f(R)$ functions in terms of the redshift $z$. We therefore highlight the basic requirements that each classes of $f(R)$ functions must satisfy and we suggest which functions better work than others.

%%%%%%%%%%%%%%%%%%%%%%%%%%%%%%%%%%%%%%
\section{Cosmography as initial settings for $f(R)$ theories}
%%%%%%%%%%%%%%%%%%%%%%%%%%%%%%%%%%%%%%

In this section, we briefly describe the principal aspects of cosmography, characterizing its implications in the context of $f(R)$ cosmology. To do so, we show how to determine numerical initial settings on the $f(R)$ function, in order to reconstruct the $f(R)$ shapes, in the interval $z\leq1$. First developments towards a complete cosmographic theory has been discussed in \cite{harry}, whereas the modern interpretation of cosmography has been firstly discussed by Weinberg \cite{weing}, who noticed that all cosmological observables may be expanded in Taylor series. Afterwards, it was soon clear that those series may be directly compared with data, without the need of postulating any cosmological model \emph{a priori} \cite{provi}. The first step is to expand the scale factor $a(t)$ around present time $t=t_0$ \cite{provi1}. Thus, the corresponding power series coefficients are referred to as the \emph{cosmographic series} (CS), if computed at $t=t_0$. Cosmography is essentially based on two simple assumptions, summarized as follows \cite{provi}:
\begin{itemize}
\item  cosmography requests the cosmological principle to hold. So, cosmography represents a model independent technique to fix cosmic bounds once spatial curvature is somehow fixed \cite{provi1}. Present-time Universe appears to be spatially flat, i.e. $k=0$, and several confirmations, coming from this assumptions, are easily determined \cite{planck}. Hence, cosmography becomes at most a model independent approach to limit cosmological models at our epoch. Since cosmography does not involve any cosmological model \emph{a priori}, it is sometimes referred to as \emph{cosmokinetics} \cite{kk};
\item current time description of our Universe passes through a direct composition of the whole energy content, by means of a non-specified number of cosmological fluids, e.g. dark energy, neutrinos, radiations, baryons, etc. Thus, the net pressure becomes a direct sum of different species, i.e. $P=\sum_i P_i$, which corresponds to separate the energy densities as: $\rho=\sum_i \rho_i$ \cite{provi1}. In our work, we circumscribe our attention to barotropic fluids only, in which the pressure becomes an energy function only.
\end{itemize}
From the above assumptions, we immediately get how to model independently define the CS. Indeed, expanding $a(t)$, we easily get
\begin{eqnarray}
a(t)  =  \sum_{n=0}^{\infty}\frac{d^n a}{dt^n}\Big|_{0}(t-t_0)^n\,,
\end{eqnarray}
and so, for our purposes:
\begin{eqnarray}\label{aditt}
{a(t)\over a_0} \sim  1 + \frac{da}{dt}\Big|_{t_0} \Delta t+ \frac{1\,}{2!}\frac{d^2a}{dt^2}\Big|_{t_0}\Delta t^2+\dots\,,
\end{eqnarray}
where we truncated $a(t)$ at the second order of Taylor expansion. For guaranteeing causality among physical processes, {\it e.g.} photon emissions from a source, we considered $\Delta t\equiv t-t_0>0$ and we finally set $a_0=1$, i.e. the scale factor today $a(t_0)$, without losing generality. The cosmographic coefficients are easily defined by:
\begin{subequations}\label{pinza}
\begin{align}
H &\equiv \frac{1}{a} \frac{da}{dt}\,,\\
q &\equiv -\frac{1}{aH^2} \frac{d^2a}{dt^2}\,,\\
j &\equiv \frac{1}{aH^3} \frac{d^3a}{dt^3}\,.
\end{align}
\end{subequations}

\begin{figure}[!hb]
        \centering
        \begin{tabular}{c}

                \begin{tabular}{c}
                   \includegraphics[width=8.5cm, height=4.5cm]{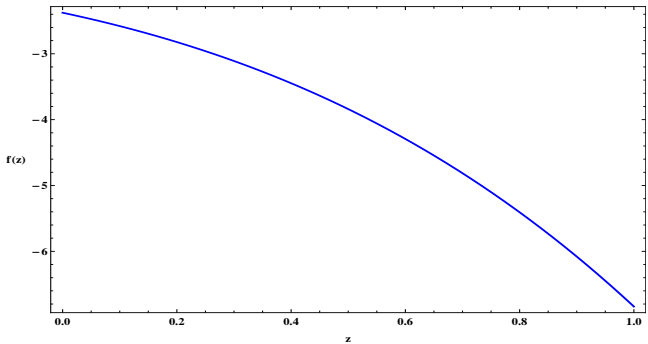}\\
                   (a) Behavior of $f(z)$ function\\[3ex]
                   \includegraphics[width=8.7cm, height=4.5cm]{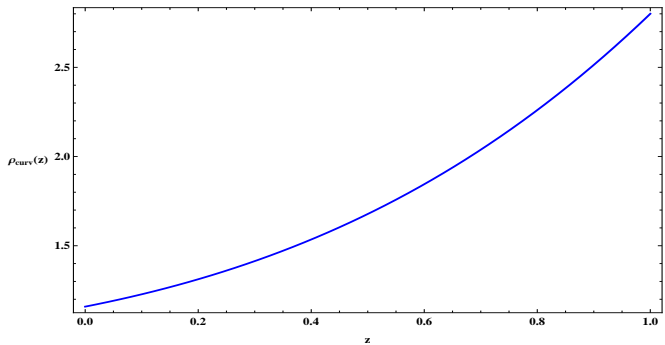}\,\,\,\,\,\\
                   (b) Behavior of $\rho_{curv}(z)$ function
                \end{tabular}

                \\

                \\
        \end{tabular}
        \caption{In these figures, we plot the functions $f(z)$ and $\rho_{curv}$ respectively in (a) and (b), in the redshift interval $z\in[0,1]$ set by assuming as the initial condition on $H(z)$ from the $\Lambda$CDM model.}
        \label{lcdm}
\end{figure}

\noindent Thus, in the light of the above definitions, the scale factor series becomes
\begin{eqnarray}\label{serie1a}
a(t) - 1\approx H_0 \Delta t - \frac{q_0}{2} H_0^2 \Delta t^2 +
\frac{j_0}{6} H_0^3 \Delta t^3\,.
\end{eqnarray}
The acceleration parameter $q$ describes whether the Universe accelerates or decelerates, whereas the jerk parameter $j$ indicates whether the acceleration changes sign in the past or continues indefinitely. For our purposes, the pressure of curvature may be numerically inferred as a source of Eq. (\ref{eq: fried2}), having in mind that the total pressure $P$ is given by $P=P_{curv} + P_m$. Assuming a perfect pressureless matter pressure, {\it i.e.} $P_m=0$, it is possible to relate the curvature pressure $P_{curv}$ to cosmography. In particular, the total pressure may be expanded as
\begin{equation}
P=\sum^{\infty}_{k=0}\frac{d^k P }{dz^k}\Big|_{t_0}z^k\,,
\end{equation}
which has been arbitrary expanded in terms of the more practical variable $z$. The pressure $P$ is intimately related to the Universe equation of state, so we need to investigate $P$ to arrive to the determination of the total equation of state, {\it i.e.} $\omega= \frac{\sum_i P_i}{  \sum_i \rho_i}$. The total equation of state of the Universe $\omega$ even contains $\omega_{curv}$ and so, its determination represents a constrain over $\omega_{curv}$ itself.

In so doing, we will be able to frame the numerical behavior of $P_{curv}$ and $\omega_{curv}$, by simply accounting the continuity equation $\frac{d\rho}{dt} + 3 H(P+\rho)=0$ and the derivatives:
\begin{subequations} \label{eq:pressureandD}
\begin{align}
P &= \frac{1}{3}H^2 \left( 2q - 1\right)\,,\label{eq:pressure} \\
\dot P & = \frac{2}{3} H^3 \left(1 - j\right)\,, \label{eq:pressure1}
\end{align}
\end{subequations}
where we made use of a more practical expression for the cosmographic coefficients, relating them to $H$ and $\dot H$, as follows \cite{provi}:
\begin{subequations}\label{eq:CSoftime}
\begin{align}
q&=-\frac{\dot{H}}{H^2} -1\,, \\
j&=\frac{\ddot{H}}{H^3}-3q-2\,.
\end{align}
\end{subequations}
It is relevant to notice that the validity of cosmographic expansions is limited to low redshift domains, i.e. $z\leq1$.  Out of this limit, Taylor series fail to converge, broadening systematics in the numerical outcomes  and so finite truncations may provide misleading results \cite{provi}.

Those issues often afflict cosmography and do not enable one to get the correct CS. Frequently, to alleviate such problems, one may add further coefficients in the Taylor expansions, while to overcome systematics one may adopt alternative parametric variables, built up as functions of the redshift $z$.  All numerical results, obtained by fixing the CS in terms of $z$ or parametric expansions indicate that \cite{provi1}
\begin{itemize}
\item the Hubble rate today seems to be underestimated by Planck results \cite{planck}, since current analyses seem to indicate a lower value. To be compatible with current cosmographic results, we take into account for our numerical analyses a normalized Hubble rate $h_0=0.68$. This value is not so relevant to determine the shape of $f(z)$ and derivatives, albeit it can enlarge or reduce the form of the curve, increasing or reducing the values of $f(z)$ and derivatives respectively;
\item the acceleration parameter $q_0$ is bounded in the interval $-0.8\leq q_0<-0.5$, indicating that the Universe is passing through an accelerated phase, different from a pure de-Sitter era \cite{ratra}. For our purposes, we employ the numerical outcome for $q_0$: $q_0=-0.57$, in order to characterize the initial conditions for $f(z)$ at our time;
\item the jerk parameter is strictly positive $j_0>0$, showing that the Universe has passed through a transition phase in which the acceleration parameter changed its sign. However, it is still not clear if $j_0\geq1$ or $j_0\leq1$. We therefore consider in our analyses the indicative value $j_0=0.77$, which apparently seems to be the most viable bound to constrain the acceleration change in the past, as suggested by recent analyses \cite{recente}.
\end{itemize}
Assuming the cited intervals, we fix the initial values for the differential equations involved in our calculations.

In the next section, we describe in detail how to relate the cosmographic recipe to the $f(R)$ function. To do so, we pass from the definition of $f(z)$, i.e. the function in which $R=R(z)$ and we set as initial values the cosmographic bounds. Moreover, we will show how to numerically solve the Friedmann equations and we highlight how to reconstruct the $f(R)$ shapes.

\begin{figure}[!hb]
        \centering
        \begin{tabular}{c}

                \begin{tabular}{c}
                   \includegraphics[width=8.5cm, height=4.5cm]{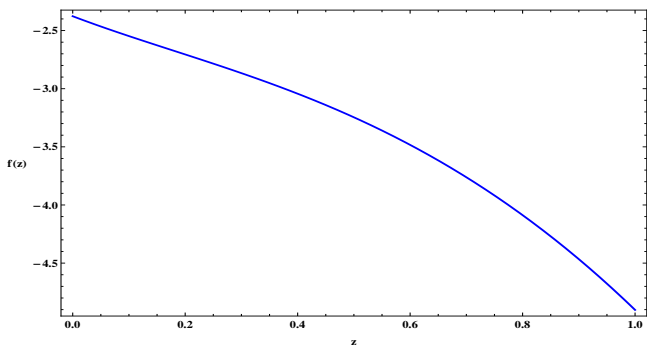}\\
                   (a) Behavior of $f(z)$ function\\[3ex]
                   \includegraphics[width=8.7cm, height=4.5cm]{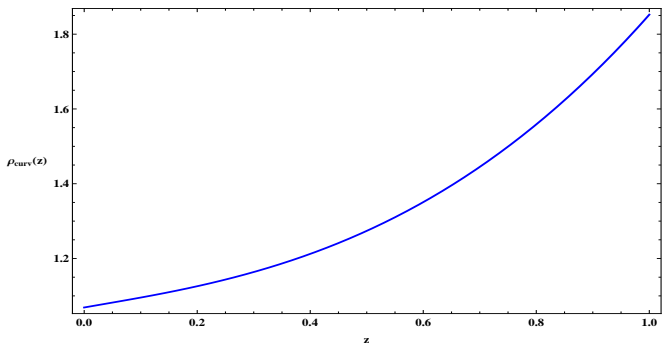}\,\,\,\,\,\\
                   (b) Behavior of $\rho_{curv}(z)$ function
                \end{tabular}

                \\

                \\
        \end{tabular}
        \caption{In these figures, we plot the functions $f(z)$ and $\rho_{curv}$ respectively in (a) and (b), in the redshift interval $z\in[0,1]$ set by assuming as the initial condition on $H(z)$ from the CPL parametrization.}
        \label{cpl}
\end{figure}

\begin{figure}[!hb]
        \centering
        \begin{tabular}{c}

                \begin{tabular}{c}
                   \includegraphics[width=8.5cm, height=4.5cm]{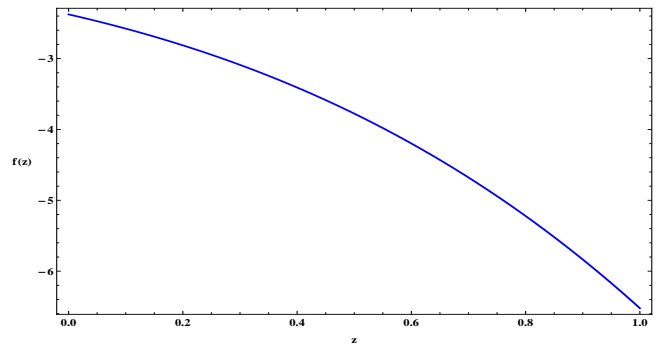}\\
                   (a) Behavior of $f(z)$ function\\[3ex]
                   \includegraphics[width=8.7cm, height=4.5cm]{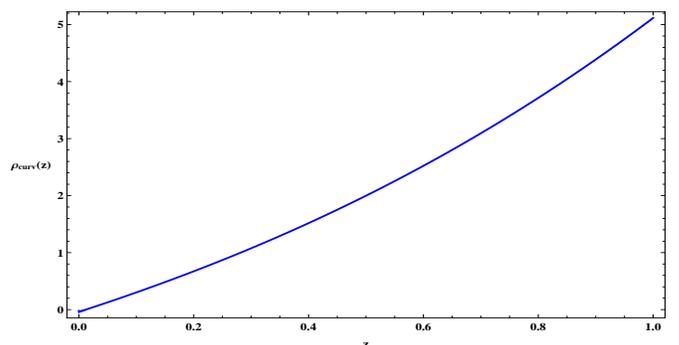}\,\,\,\,\,\\
                   (b) Behavior of $\rho_{curv}(z)$ function
                \end{tabular}

                \\

                \\
        \end{tabular}
        \caption{In these figures, we plot the functions $f(z)$ and $\rho_{curv}$ respectively in (a) and (b), in the redshift interval $z\in[0,1]$ set by assuming as the initial condition on $H(z)$ from the phenomenological approach.}
        \label{staro}
\end{figure}

%%%%%%%%%%%%%%%%%%%%%%%%%%%%%%%%%%%%%%%%%%%%%%%%%%%%%%%%
\section{Cosmographic reconstructions of $f(R)$ functions}
%%%%%%%%%%%%%%%%%%%%%%%%%%%%%%%%%%%%%%%%%%%%%%%%%%%%%%%%

The possibility to make a correspondence between modified theories of gravity and cosmography has reached a reasonable point, since it is now possible to relate the cosmographic series to $f(R)$, passing through the determination of the corresponding $f(z)$ function. Rephrasing it differently, one may obtain experimental constraints on $f(R)$ models trough cosmography, selecting which model is effectively compatible with late- time bounds. The first step is to rewrite $f(R)$ in function of the redshift $z$, thence obtaining  $f(R)\rightarrow f(R(z))\equiv f(z)$. The function $f(z)$ represents the $f(R)$ function, evolving in the redshift domain. It is more practical to  handle $f(z)$ than $f(R)$ for our numerical approaches, because $f(z)$ may be directly compared with the Universe evolution in terms of the redshift. Hence, we first manage to obtain a possible class of $f(z)$ and then getting back, we infer the corresponding $f(R)$. The reason of using $z$ instead of $R$ lies on the fact that all observable quantities, entering the Friedmann equations, can be easily framed in terms of $z$. Thus, the corresponding differential equation one gets is function of the redshift only, and can be numerically solved.

For our purposes, we first express the Ricci scalar in terms of $z$ and $H(z)$. To determine the Ricci evolution, one has to somehow characterize the Hubble parameter $H(z)$, i.e.
\begin{equation}\label{18}
R=6[(1+z) HH_z - 2H^2]\,,
\end{equation}
where we substitute the time derivative  in (\ref{eq: constr}) with this expression: $\frac{d}{dt}=-(1+z) H(z) \frac{d}{dz}$.
Please note that hereafter for any function $X(x)$, depending upon the auxiliary variable $x$, we define the $i$th derivative as
\begin{equation}\label{skh}
X_{ix}=\frac{d^iX}{dx^i}\,,
\end{equation}
and, if evaluated at $t=t_0$, as $X_{iz0}=\frac{d^iX}{dx^i}\Big|_0$ \cite{provi}. Thus, in Eq. (\ref{18}) $H_z$ denotes the first derivative with respect to the redshift $z$.

We need to solve the following dynamical problem:
\begin{eqnarray}\label{ldjfkeh}
H&=&H(\rho, f(R), f^\prime(R),\dot H)\,,\nonumber\\
R&=&R(t)\,,\\
H&=&H(t)\,,\nonumber
\end{eqnarray}
and by using $a(z)=(1+z)^{-1}$, we can solve the dynamical problem, by numerically framing Eq. (\ref{eq: fried1}) by means of a single variable, i.e. the redshift $z$. To do so, we need to determine the time dependence with respect to the redshift and how the curvature evolves as the redshift varies. Hence,
\begin{equation}\label{orl}
\frac{dz}{dt}=-(1+z)H(z)\,,
\end{equation}
and simply having
\begin{equation}\label{ucazz}
d\ln\left(z+1\right)=3\frac{dH^{2}}{R+12H^{2}}\,,
\end{equation}
we finally obtain
\begin{equation}\label{mia1}
z\left(R\right)=z_{0}\exp\left\{\int
\frac{3dH^{2}}{R+12H^{2}}\right\}-1,
\end{equation}
which represents the expression permitting $z$ to evolve in terms of $R$.

Thus, considering $H=H(z)$, it naturally follows $R=R(z)$. Moreover, we write down the following expressions
\begin{equation}\label{Rinz0}
\begin{split}
\frac{R_0}{6} =\, & H_0\left[H_{z0}-2H_0\right]\,,\\
\frac{R_{z0}}{6} =\,& H_{z0}^2+H_0(-3H_{z0}+H_{2z0})\,,\\
\end{split}
\end{equation}
which relate the Ricci scalar today and its first derivative to the CS and that we obtain considering Eq. (\ref{18}) and Eq. (\ref{orl}).

Our aim is to feature the shape of Universe's dynamics in the observable interval $z\in[0,1]$. To do so, i.e. to arrive to reconstruct $f(R)$ in a model independent way, we need to impose how $H(z)$ evolves in such an interval. In fact, considering the first Friedmann equation (\ref{eq: fried1}), and rewriting it in terms of the redshift $z$, it is clear that, by means of the cosmographic results, adopted in the context of $f(R)$ gravity, we need to allow $H(z)$ to evolve in the observable limit $z\in[0,1]$. To do this, our strategy is employing different Hubble rates, based on well-known paradigms, capable of describing Universe's dynamics at small redshifts and reproducing in sequence $f(z)$ and $f(R)$. In particular, we consider the Hubble rates derived in the $\Lambda$CDM model, in the CPL parametrization, and in a phenomenological reconstruction in powers of $\propto a(t)$ of dark energy. We therefore proceed as reported in the next: we separately assume the three $H(z)$ expressions and rewrite the Friedmann equation (\ref{eq: fried1}) in terms of the redshift parameter, by using the expressions (\ref{orl}, \ref{Rinz0}). As numerical initial conditions, we assume the validity of numerical bounds inferred from cosmography. In particular, we find $f_0\equiv f(z=0)$ and $f_{z0}\equiv f'(z=0)$ from (\ref{f0fz0fzz0dopo}) with the constraints
\begin{subequations}\label{indiao}
\begin{align}
f^\prime(R_0)&=1\,,\\
f^{''}(R_0)&=0\,,
\end{align}
\end{subequations}
which respectively indicate that at the solar system level, the gravitational constant $G$ acts as observations indicate, without any departure and general relativity is easily recovered as  $f(R)\rightarrow R=R_0$. In addition, we employ for $q_0$ and $j_0$ the values reported in Sec. III.

Then we solve the differential equation (\ref{eq: fried1}), replacing $f(R), f'(R), f''(R)$ in functions of a single variable, i.e. the redshift $z$. Afterwards, passing from the time derivative to redshift derivative, by means of (\ref{mia1}). To express the Hubble evolution in terms of the redshift $z$, we take into account the following expressions \cite{provi}
\begin{equation}\label{Hpunto}
\begin{split}
\frac{dH}{dt} =& -H^2 (1 + q)\,,\\
\frac{d^{2}H}{dt^2} =& H^3 (j + 3q + 2)\,,\\
\end{split}
\end{equation}
where rewriting all in function $z$ (by means of Eq. (\ref{orl})),  we can find useful constraints for Hubble rate and its derivative in function of cosmographic set at $z=0$:
\begin{equation}\label{Hpunto}
\begin{split}
\frac{H_{z0}}{H_0}=& 1+q_0  ,\\
\frac{H_{2z0}}{H_0}=& j_0-q_0^2 \,.\\
\end{split}
\end{equation}
 By means of (\ref{Rinz0}) and (\ref{Hpunto}), we evaluate $R$ in function of the cosmographic series only, i.e.:
 
 \begin{subequations}\label{ricciederricci}
\begin{align}
R&=6H_0^2(q_0-1),\\
R_z&=6H_0^2(-2-q_0+j_0),
\end{align}
\end{subequations}
where $\Omega_m$ is the matter density.
 
  So that, we can rewrite $f(R)$ derivative in function of $R$ as derivative in function of $z$. It is straightforward to show  that  the final result is:
\begin{equation}\label{zuzzu}
\begin{split}
f'(R) =\, & \frac{f_z}{R_z}\,,\\
f''(R)=\, & \frac{(f_{2z}R_z - f_zR_{2z})}{R_z^{3}}\,,\\
f'''(\mathcal{R})=\, &\frac{f_{3z}}{R_z^3} - \frac{f_z\, R_{3z}+3f_{2z}\,
R_{2z}}{R_z^4}+\frac{3f_z\, R_{2z}^2}{R_z^5}\,.
\end{split}
\end{equation}

Besides expanding $f(z)$, in terms of the cosmographic parameters, we have \cite{provi}
\begin{equation}\label{f0fz0fzz0dopo}
\begin{split}
\frac{f_0}{2H_0^2}=\,&-2+q_0\,,\\
\frac{f_{z0}}{6H_0^2} =\,&-2-q_0+j_0\,,\\
\frac{f_{2z0}}{6H_0^2}=\,&-2-4q_0-(2+q_0)j_0-s_0\,.\\
\end{split}
\end{equation}

Straightforwardly, substituting the expressions of $R(z)$ in function of CS (\ref{ricciederricci}) and expressions of $f(z)$ in terms of CS (\ref{f0fz0fzz0dopo}) in (\ref{zuzzu}), we have the derivatives of $f$ in function of CS  at present time. Those would represent the initial settings, determined from cosmography on $f_0 $ that we will use as initial conditions to solve Eq. (\ref{eq: fried1}) in the case of values of $H_0$, $q_0$ and $j_0$ indicated in Sec. III.

Nevertheless, since $R(z)$ is a  invertible function, we simply compute $f(R)$, numerically solving the following integral
\begin{equation} \label{integrale}
f(R)= \int{\frac{df}{dR}dR}=\int{\frac{df}{dz}\frac{dz}{dR}dR}+K_{cs}\,,
\end{equation}
where we substituted the expression of $z(R)$ obtained matching (\ref{eq: constr}) and the related expression for $H(z)$. The integration constant, $K_{cs}$, is determined from the cosmographic results, giving for example in the $\Lambda$CDM case the approximate value $K_{cs}\approx 20.9\,R_0$. It is necessary to notice that $K_{cs}$ is not related to the cosmological constant. Trough another procedure we can determinate $K_{cs}$  estimating the difference between $f(R)$ found by integration as in (\ref{integrale}) and $f(R)$ found putting $z(R)$ into $f(z)$. Both the procedures permit to infer a class of $f(R)$ functions, from assuming $f(z)$, but they differ from the constant $K_{cs}$. The increasing or decreasing ratio between the two procedures is the constant $K_{cs}$, which comes from the definite integration that we perform.

\begin{figure}[!hb]
        \centering
        \begin{tabular}{c}

                \begin{tabular}{c}
                   \includegraphics[width=8.5cm, height=4.5cm]{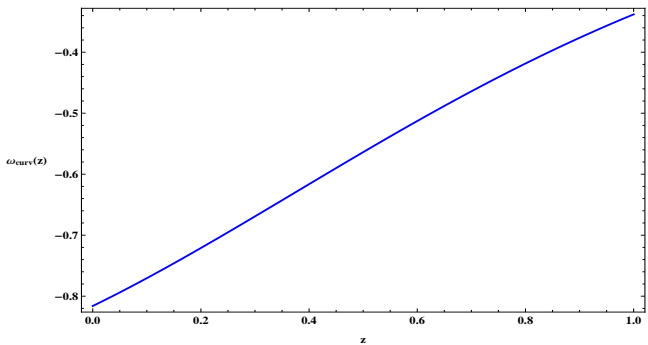}\\
                   (a) Behavior of $w_{curv}$ in the $\Lambda$CDM case\\[3ex]
                   \includegraphics[width=8.7cm, height=4.5cm]{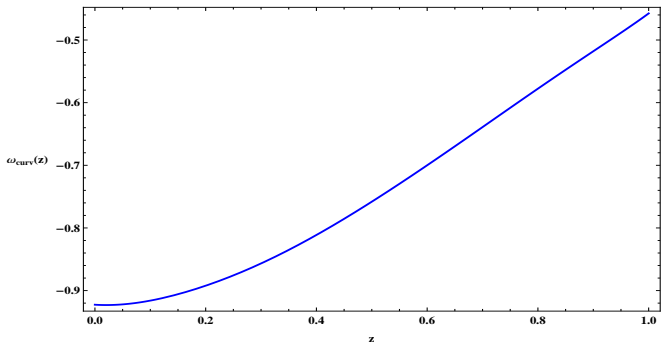}\,\,\,\,\,\\
                   (b) Behavior of $w_{curv}$ in the CPL case\\[3ex]
                   \includegraphics[width=8.7cm, height=4.5cm]{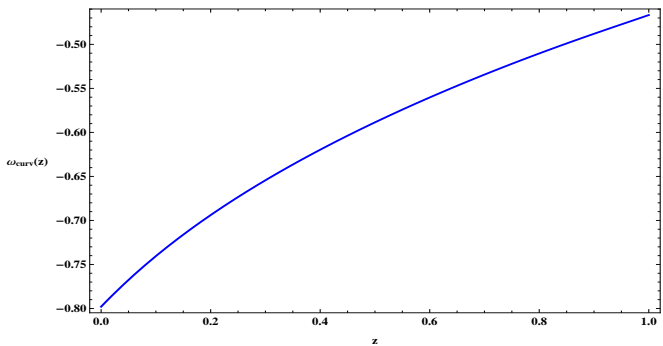}\,\,\,\,\,\\
                   (c) Behavior of $w_{curv}$ in the phenomenological case
                                   \end{tabular}

                \\

                \\
        \end{tabular}
        \caption{Here, we plot the behaviors of different $w_{curv}$ for the cases of the $\Lambda$CDM model, CPL parametrization and phenomenological reconstruction.}
        \label{DMpioneers}
\end{figure}

\begin{figure}[!hb]
        \centering
        \begin{tabular}{c}

                \begin{tabular}{c}
                   \includegraphics[width=8.5cm, height=4.5cm]{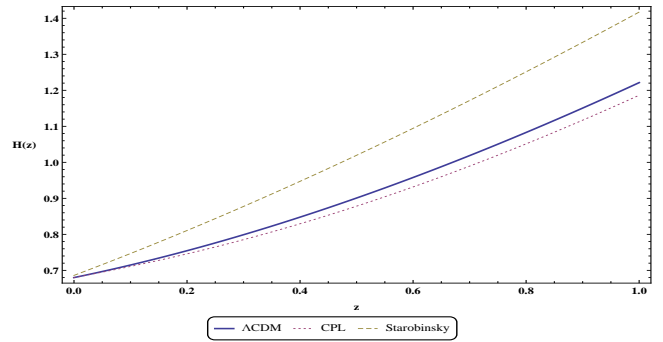}\\
                   (a) Behaviors of different $H(z)$\\[3ex]
                   \includegraphics[width=8.7cm, height=4.5cm]{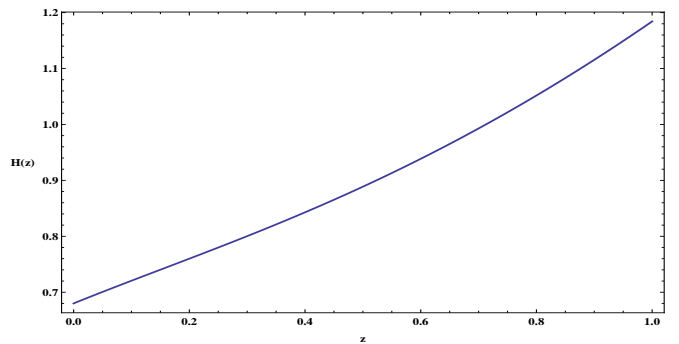}\,\,\,\,\,\\
                   (b) Behavior of our numerical $H_{curv}$
                \end{tabular}

                \\

                \\
        \end{tabular}
        \caption{In these figures the different Hubble rates for the $\Lambda$CDM model, CPL parametrization and phenomenological reconstruction are reported, in the redshift interval $z\in[0,1]$. Besides, we also report the Hubble rate evaluated the general $f(R)$ obtained in our analysis.}
        \label{Hubbleratee}
\end{figure}

\begin{figure}[!hb]
        \centering

                   \includegraphics[width=8.7cm, height=4.5cm]{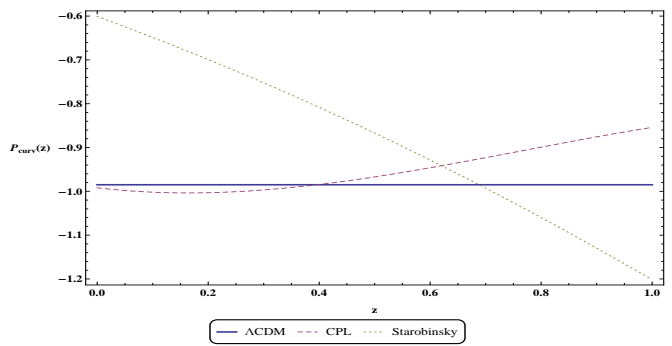}

        \caption{Behavior of our numerical $P_{curv}$
                }
        \label{DMpioneers1}
\end{figure}

Please note that the determination of $f(z)$ and of the related  $f(R)$ passes  through defining the Hubble rate in the observable interval $z\in[0,1]$  that we will use to solve Eq. (\ref{eq: fried1}).
\\
 In our computation we adopt three cosmological setting Hubble rates. Every model assumes dark matter and baryons to evolve separately from dark energy. The models are itemized hereafter.
 \begin{itemize}
 \item In the first case, i.e. the $\Lambda$CDM model, the dark energy density $\Omega_X$ is constant at every epochs of Universe's expansion, providing $\Omega_X\equiv1-\Omega_m$, where $\Omega _m$ is the matter density.
 
 \item In the second case, CPL \cite{provi1, Polar}, the dark energy equation of state is expanded around $a=1$, at a first order of Taylor expansion, giving $\omega=\omega_0+\omega_1 (1-a)$ \cite{Polar}. The dark energy term is proportional to $\Omega_X=(1-\Omega_m)(1+z)^{3(1+\omega_0+\omega_1)}e^{-\frac{3 \omega_1 z}{1+z}}$. 

\item Finally, the third approach involves a more phenomenological framework in which, in addition to standard matter, one includes $\Omega_X=\text{A}_1+\text{A}_2 (1+z)+\text{A}_3 (1+z)^2$, corresponding to a some sort of Taylor expansion around $a^{-1}$, truncated at the second order \cite{staro}, with $A_1=0.2$, $A_2=0.2$, $A_3=0.3$, three phenomenological constants.
\end{itemize}
The graphic representation  of the three different behaviors of Hubble rates is in Fig. \ref{Hubbleratee}.
 For each model, we compute $f(z)$ and the associated $f(R)$, together with $P_{curv}$ and $\rho_{curv}$ for $z$ spanning the interval from $0$ to $1$. The corresponding expressions for $\rho_{curv}$ and $\omega_{curv}$ in terms of the redshift $z$ are reported in the Appendix A. The numerical results for  $\rho_{curv}$ and $P_{curv}$ have been reported in Figs. \ref{DMpioneers} and Fig. \ref{DMpioneers1}.
 The behavior of $f(z)$ and $\rho_{curv }$ is reported in Figs. \ref{lcdm}, \ref{cpl}, \ref{staro}.

%%%%%%%%%%%%%%%%%%%%%%%%%%%%%%%%%%%%%%%%%%%%%%%%%%%%%%
\section{Analysis of cosmographic results}
%%%%%%%%%%%%%%%%%%%%%%%%%%%%%%%%%%%%%%%%%%%%%%%%%%%%%%

We get three expressions for $f(z)$ and $f(R)$, for a total of six analyses. All the numerical results, performed by using the three Hubble rates, i.e. $\Lambda$CDM, CPL, and the phenomenological one, generate close outcomes which converge to the approximate solution
\begin{equation}\label{fdir}
\begin{split}
f_{eff}(R)= R+\Lambda -2 a (R-R_0)-a (R-R_0)^2+\\-5\: b\: e^{b R} ArcTang(R-R_0)-c\: Sin(R-R_0),
\end{split}
\end{equation}
which represents an effective form of $f(R)$ inferred from reconstructing $f(R)$ from our numerical curves. For guaranteeing that Eq. (\ref{fdir}) accurately reproduces the Universe dynamics, we set the three free constant around the following values:
\begin{subequations}\label{costanti}
\begin{align}
a&=0.021\\
b&=22.4\\
c&=0.0098,
\end{align}
\end{subequations}
while $ \Lambda $ represents the cosmological constant and $R_0$
is the present value of the Ricci scalar, which becomes $R_0\approx-4.4$ using the cosmographic results. Moreover, Eq. (\ref{fdir}) accurately reproduces the gravity lagrangian in the range  $R \in [-15,-4.4]$. All curves of $f(z)$ and $f(R)$, approximating to $R_0=-4.4$, provide the relevant fact that Eq. (\ref{fdir}) reduces to the Hilbert-Einstein action  $f(R)= R + \Lambda$, recovering general relativity with the addition of a cosmological constant. Rephrasing it differently, the $\Lambda$CDM model persists to be favorite at small redshift domains, indicating however the cosmological constant as a limiting case of a more general paradigm. Another significative result is offered by reconstructing the effective Hubble rate from the first Friedmann equation Eq. (\ref{eq: fried1}), as due to the $f(R) $  corrections of Eq. (\ref{fdir}). We find, in particular, that a viable approximation is the one plotted in Fig. (\ref{DMpioneers}). The approximation may be framed in terms of
\begin{equation} \label{H}
H^2(z)=H_0^2\left(\Omega_m(z+1)^3+\delta  \tanh (\alpha +\beta  z)+\tau  e^{\gamma  z}\right)\,,
\end{equation}
where
\begin{subequations}\label{coeffffff}
\begin{align}
\alpha &=2.43\,,\\
\beta &=3.35\,, \\
\tau &= -6.74\,,\\
\delta &=7.54\,,\\
\gamma &=0.046\,,
\end{align}
\end{subequations}
that are in agreement with experimental cosmological prediction. Here, $\Omega_m$ is fixed in terms of the Planck results \cite{plankkk}. The behaviors of $\rho_{curv}$ and $\omega_{curv}$ provide similar contributions in all the three models investigated and their numerical difference are proportional to the percent difference settled by the three different Hubble rate. This suggests that our numerical outcomes do not strongly depend upon the initial conditions on $H$ and on its evolution.

%%%%%%%%%%%%%%%%%%%%%%%%%%%%%%%%%%%%%%%%%%%%%%%%%%%%%%
\section{Final outlooks and perspectives}
%%%%%%%%%%%%%%%%%%%%%%%%%%%%%%%%%%%%%%%%%%%%%%%%%%%%%%

In this paper, we described a method to frame the correct $f(R)$, basing our attention on numerically solving the modified Friedmann equations. Among several possibilities to feature the $f(R)$ by postulating its shape according to observations, we propose a new technique consisting in assuming initial settings on the modified Friedmann equations offered by cosmography. The procedure shows the advantage to be model independently fixed in the redshift interval $z\in[0,1]$, leading to a $f(R)$ reconstruction which well fits the experimental outcomes provided by present data. Thus, we managed to alleviate the degeneracy among the different choices of $f(R)$ functions, by initially reconstructing the dynamic equations in terms of the single variable $z$. Indeed, employing the correspondence $R=R(z)$ and showing its invertibility in the redshift range $[0,1]$, one can shift all variables in terms of the redshift $z$, showing that all observable quantities of interest can be framed analogously. The differential equation we got has been derived by featuring the first Friedmann equation in terms of $z$. The initial conditions on its dynamics have been bounded by means of cosmography, i.e. a numerical technique which determines cosmic bounds at late times, by simply comparing Taylor expansions directly to data. The powerful of cosmography, i.e. the fact of being completely model independent, set up the initial conditions on the $f(R)$ functions and enables one to reproduce Universe's expansion history as the knowledge of $H(z)$ is somehow known. We proposed three Hubble rates which actually provide the numerics in the interval $z\leq1$, i.e. the $\Lambda$CDM, the CPL and the phenomenological approach. We aimed to choose those frameworks since are likely the simplest frameworks which better adapt their shapes to frame the Universe evolution. Correspondingly, we got a model independent reconstruction of $f(z)$, i.e. the function $f(R)$ in terms of the redshift $z$. We reported the shapes of $f(z)$ for each model involved in our numerical analyses and we showed that our outcomes are basically similar but not equivalent, obtaining consequently different classes of plausible $f(z)$ functions. In so doing, we passed through a numerical description of $f(R)$ functions, by means of the auxiliary $f(z)$. In fact, we got a defined class of viable functions which integrate the numeric Friedmann equations, providing integration constants accurately fine-tuned by means of cosmography itself. The final outcome lies on  viable forms of  $f(R)$ functions, which extend the standard approach of general relativity. The terms here involved are compatible with recent developments on modified $f(R)$ functions and permit  to conclude that the Universe is accurately featured by $R$ corrections to the standard Hilbert-Einstein action.
Besides, in order   to extend $f(R)$ effective expression till to the dynamics of the early Universe, we are trying to apply our numerical procedure, here described, for redshift parameter major than unity, i.e. $z\geq 1$. This topic will be the subject of a future work.

%%%%%%%%%%%%%%%%%%%%%%%%%%%%%%%%%%%%%%%%%%%%%%%%%%%%%%
\section*{Acknowledgements}
%%%%%%%%%%%%%%%%%%%%%%%%%%%%%%%%%%%%%%%%%%%%%%%%%%%%%%

 The author wishes to express his gratitude to G. Lambiase for discussions on the topic of present work and A. Strumia for his support. The author is finantially by INFN section of Pisa. A particular thanks to E. Vicari for his support and useful discussions.

%%%%%%%%%%%%%%%%%%%%%%%%%%%%%%%%%%%%%%%%%%%%%%%%%%%%%%

%%%%%%%%%%%%%%%%%%%%%%%%%%%%%%%%%%%%%%%%%%%%%%%%%%%%%%

\begin{widetext}

%%%%%%%%%%%%%%%%%%%%%%%%%%%%%%%%%%%%%%%%%%%%%%%%%%%%%%
\appendix
%%%%%%%%%%%%%%%%%%%%%%%%%%%%%%%%%%%%%%%%%%%%%%%%%%%%%%
\newpage

\textbf{Appendix A}
\\
In this appendix we show the analytical expression for the barotropic density  in function of z, i.e.
\begin{equation}
\begin{split}
\rho_{curv}(z)= \frac{1}{2} \left\lbrace  \frac{f(z)}{f'(z)}R'(z) -R \right\rbrace -3 H^2 (1+z)\left\lbrace R''(z)\frac{1}{R'(z)}- \frac{f''(z)}{f'(z)}\right\rbrace ,
\end{split}
\end{equation}
and  barotropic curvature factor $\omega_{curv}$ in function of z:
\begin{equation}\label{eq: baroz}
\begin{split}
 \omega_{curv}=-1+\frac{B}{A}
\end{split}
\end{equation}
where
\begin{equation}
A=R'(z) \left(R'(z) \left(6 H(z) \left((z+1) H(z) f''(z)+f'(z) \left(2 H(z)-(z+1) H'(z)\right)\right)+f(z) R'(z)\right)-6 (z+1) H(z)^2 f'(z) R''(z)\right),
\end{equation}
and
\begin{gather}
B =  2 (z+1) H(z)\cdot \notag \\ \cdot \Bigg[(z+1) H'(z) R'(z)  \Big(f''(z) R'(z)-f'(z) R''(z)\Big)+H(z) \Bigg(2 (z+1) f'(z) R''(z)^2+\Big((z+1) f^{(3)}(z)+2 f''(z)\Big) R'(z)^2+ \notag \\+R'(z)  \Big(-(z+1) R^{(3)}(z) f'(z)-2 R''(z) \Big((z+1) f''(z)+f'(z)\Big)\Big)\Bigg)\Bigg].
\end{gather}
These expressions have been used in this manuscript to compute the correct expression of $\rho_{curv}  $ and $\omega_{curv}$ from $f(z)$.
\end{widetext}

\end{document}